\documentclass[12pt]{iopart}

\usepackage{dcolumn}% Align table columns on decimal point
\usepackage{bm}% bold math
\usepackage{graphicx}% Include figure files
\usepackage{color}
\usepackage{subfigure}
\usepackage{hyperref}
\usepackage{latexsym}
\usepackage{amsthm}
\usepackage{amssymb}
\usepackage{braket}

\usepackage{cite}
\DeclareGraphicsExtensions{.jpg,.pdf, .mps, .png, .eps, .ps, .EPS,.gif}

\newcommand{\Avg}[1]{\left\langle{#1}\right\rangle}
\newcommand{\dirac}{\partial \hskip -2.0mm \slash} 

\newcommand{\A}{A \hskip -2.0mm \slash} 
\newcommand{\B}{B \hskip -2.0mm \slash} 

\newcommand{\iu}{\mathrm{i}}
\begin{document}
\def\be{\begin{equation}}
\def\ee{\end{equation}}

\def\bc{\begin{center}}
\def\ec{\end{center}}
\def\bea{\begin{eqnarray}}
\def\eea{\end{eqnarray}}

\def\ie{\textit{i.e.}}
\def\etal{\textit{et al.}}
\def\m{\vec{m}}
\def\G{\mathcal{G}}

\newcommand{\gin}[1]{{\bf\color{cyan}#1}}

\title[Dirac gauge theory for topological spinors in 3+1 dimensional networks]{Dirac gauge theory for topological spinors in 3+1 dimensional networks}

\author{Ginestra Bianconi$^{1,2}$}

\address{$^1$ School of Mathematical Sciences, Queen Mary University of London, E1 4NS London, United Kingdom\\
$^2$ Alan Turing Institute, The British Library, London, United Kingdom}

\ead{ginestra.bianconi@gmail.com}
\vspace{10pt}
\begin{indented}
\item[]
\end{indented}

\begin{abstract}
Gauge theories on graphs and networks are attracting increasing attention not only as approaches to quantum gravity but also as models for performing quantum computation.
Here we propose a Dirac gauge theory for topological spinors in $3+1$ dimensional networks associated to an arbitrary metric. Topological spinors are the direct sum of  $0$-cochains and $1$-cochains defined on a network and describe a matter field defined on both nodes and links of a network. Recently in Ref. \cite{bianconi2021topological} it has been shown that topological spinors obey the topological  Dirac equation driven by the discrete Dirac operator. 
In this work we extend these  results by formulating the Dirac equation on weighted and directed $3+1$ dimensional networks  which allow  for the treatment of  a local theory. The commutators and anti-commutators of the Dirac operators are non vanishing an they define the curvature tensor and magnetic field of our theory respectively. This interpretation is confirmed by the non-relativistic limit of the proposed Dirac equation. In the non-relativistic limit of the proposed Dirac equation  the sector of the spinor defined on links follows the Schr\"odinger equation with the correct giromagnetic moment, while the sector of the spinor defined on nodes follows the Klein-Gordon equation and is not negligible. The action associated to the proposed field theory  comprises of  a  Dirac action  and a metric action. 
We describe the gauge invariance of the action under both Abelian and non-Abelian transformations and we propose the equation of motion of the  field theory of both Dirac and metric fields. This theory can be interpreted as a limiting case of a more general gauge theory valid on any arbitrary network in the limit of  almost flat spaces.
\end{abstract}

%\maketitle

\section{Introduction}
In this work we formulate a Dirac gauge theory on  networks in which the matter field is defined on both nodes of links of the network extending previous work on the topological Dirac equation \cite{bianconi2021topological}.  
Recently there is a growing attention on lattice gauge theories \cite{rothe2012lattice} and their implementation using quantum information frameworks \cite{dalmonte2016lattice,banuls2020simulating,aidelsburger2018artificial,surace2020lattice}.
While in lattice gauge theories  gauge fields are often associated to the links of a network, matter fields are traditionally only associated  to the nodes of the network  \cite{tagliacozzo2013simulation}. Interestingly however, models in which  quantum states are defined on links and on higher-dimensional simplices or cells of  simplicial or cell complexes are receiving growing interest and they include for instance the very influential Kitaev model \cite{kitaev2003fault}. 
Also in the context of classical dynamics  on networks it is becoming clear that considering dynamical variables not only associated to the nodes but also to the links of networks and to the higher-dimensional simplices of simplicial complexes can reveal the important interplay between network topology and dynamics \cite{bianconi2021higher,millan2020explosive,battiston2021physics,carletti2022global}.
In this context it is emerging that the discrete Dirac  operator\cite{bianconi2021topological,post2009first,davies1993analysis,
requardt2002dirac,hinz2013dirac,lloyd2016quantum} is key to investigate the properties of coupled topological signals of different dimension defined on graphs and networks \cite{calmon2022dirac,giambagli2022diffusion,calmon2022local}. 

The discrete Dirac operator on networks is a  direct extension of the Dirac operator used in the continuum and its  numerical implementations  \cite{fillion2013formal,flouris2022curvature}. The Dirac operator plays an important role for example in supersymmetric theories \cite{witten1982supersymmetry}  and  in non-commutative geometry  \cite{connes2019noncommutative,beggs2020quantum,lira2022geometric,cipriani2014spectral,paschke1998discrete,krajewski1998classification} where to our knowledge it has been first formulated \cite{davies1993analysis} [Note however that that this {\em topological } definition of the Dirac operator is distinct from the one used in quantum graphs literature where links are considered linear one dimensional spaces on which a $d+1=1+1$ Dirac equation is defined \cite{bolte2003spectral,bolte2003spin,fijavvz2021linear,kramar2020linear,kramar2021dynamic}.] The Dirac operator  can be defined on networks also without relying on non-commutative geometry \cite{bianconi2021topological,post2009first,lloyd2016quantum,hinz2013dirac,anne2015gauss,athmouni2021magnetic,requardt2002dirac,miranda2023spectral,parra2017spectral,miranda2023continuum}.
However most of the works that use the discrete Dirac operator outside the field of noncommutative geometry define  the discrete Dirac operator on a network as  the Hodge-Dirac operator (also called Gauss-Bonnet operator) given by the sum of the exterior derivative and its dual $\partial=d+d^{\star}$, i.e. they do not   associate an algebra to these operators.

The Dirac equation proposed in Ref. \cite{bianconi2021topological} is based on a topological description of the wave function formed by a topological spinor defined on nodes and links of a network. In particular Ref. \cite{bianconi2021topological} considers the topology of finite dimensional $1+1,2+1 $ and $3+1$ square lattices and associates an algebra to the Dirac operators defined on space-like and time-like directions as well. 

In this work we build on these results to formulate a gauge theory for a Dirac field defined on both nodes and links of a network and its associated metric degrees of freedom.
 The formulated gauge theory framework can be considered  as a  gauge theory which could be potentially explored for synthetic realization in condensed matter systems. Alternatively the proposed gauge theory can be considered in the framework of the current scientific debate on the topic in the context of quantum gravity
 \cite{Oriti,baez1996spin,matsuura2022supersymmetric,jiang2022gauge}. 

From a mathematical perspective quantum gravity is about reconciling quantum mechanics, a theory that does not have a fundamental geometrical foundation but is instead characterized by its probabilistic interpretation with general relativity, that is a deeply rooted in a geometrical description of space-time.
Many approaches \cite{CDT,Rovelli,calcagni2013laplacians,
Lionni,Dario,Astrid,Codello} to quantum gravity aim at turning general relativity into a quantum discipline, however one different strategy is to try to turn quantum mechanics into a geometrical and topological theory leading to classic approaches and results including \cite{nakahara2003geometry,witten1982supersymmetry}.
In this respect, with  the flourishing of results on graphs and networks, in the last years we have seen a growing interest in  approaches combining graph and network theory to quantum gravity and viceversa combining quantum gravity approaches to networks
 \cite{bianconi2021higher,Astridnet,reitz2020higher,Severini,Trugenberger1,Trugenberger2,kleftogiannis2022physics}
and addressing questions related to  emergent geometry \cite{bianconi2017emergent,bianconi2016network,wu2015emergent,chen2013statistical,akara2021birth,kleftogiannis2022emergent} and quantum information \cite{anand2011shannon,biamonte2019complex,de2016spectral,bottcher2022complex,de2015structural,nokkala2020probing,nokkala2016complex,nokkala2018reconfigurable}.

In this perspective of course it seems natural to rely on algebraic topology, exterior calculus, and discrete geometry \cite{eckmann1944harmonische,horak2013spectra,calcagni2013laplacians,desbrun2005discrete,grady2010discrete,lim2020hodge,Tee} and start by interpreting geometrically and in the discrete network setting, the relativistic  Dirac equation  \cite{dirac} which historically has stimulated fundamental research in both physics and mathematics \cite{thaller2013dirac,pais2005paul}.
Driven by this abstract line of thought,  the topological Dirac equation \cite{bianconi2021topological} has been recently proposed.
The Dirac equation defined in Ref. \cite{bianconi2021topological} acts on a topological spinor  defined on nodes and links, in which the wave function is slightly non-local as on each link the wave function takes the same value.
Here we show that the Dirac equation \cite{bianconi2021topological} can be also defined on a {\it local} wave function defined on the  nodes and on each directed link departing from every node $i$ of the network, thus acting as the fiber bundle at node $i$. Therefore the topological spinor can be interpreted as encoding both the component of the wave-function defined on each  node of the network and a "flux" of the wave function from a given node toward any neighbour node. Therefore the wave-function has an interpretation that is  similar to the description of a dynamical state of a classical particle requiring both position and velocity of the particle.
Here not only we turn the topological Dirac equation into a local theory, but we also  introduce metric matrices on nodes and links.

The metric matrices inducing deformations of the lattice are here  treated as the geometric degrees of freedom of the network while the topology remains the $3+1$ dimensional one. The  metric degree of freedoms are embodied in a metric field that  evolves together with the topological spinor according to the equations of motion. 

Building on the fact that the directional Dirac operators along different directions of the lattice do not commute or anticommute, we define a curvature tensor  and the magnetic field of the network, from the commutators and anticommutators of the directional Dirac operators.

The non-relativistic limit of the proposed Dirac equation   confirms our interpretation of the anti-commutators of the spatial Dirac operators as the magnetic field of the network. Therefore in this context the electromagnetic field of this theory is associated to the metric of  the network.
Moreover from the non-relativistic limit of the Dirac equation we learn the following results.
First, the  Schr\"odinger equation with the correct giromagnetic factor is recovered for the sector of the spinor associated to the links (or the fiber bundle).
Secondly, the node sector of the topological spinor follows the Klein-Gordon equation.
Thirdly, the node sector of the topological spinor is non-negligible.
It is interesting to consider whether these results can lead  to testable predictions different from the standard Dirac equation. 
%More theoretically these results open new perspectives for connections to supersymmetry, in particular when considering the second quantization of the topological spinor a question that emerges if whether $\bm\chi$.

Finally we define the Abelian and non-Abelian transformations acting simultaneously on the topological spinors and the network metric, and we derive the equation of motion from our action which relates the metric degree of freedom to the topological spinor.

The proposed gauge field theory on $3+1$ dimensional networks can be interpreted as a limiting case of a more general theory valid on an arbitrary networks which applies when the discrete space-time is almost flat.  The extension to arbitrary network topologies is not excluded but it would require changing the algebra adopted for the directional Dirac operators.

\section{Dirac equation for topological spinors}
We consider a lattice in $3+1$ dimensions with $N$ nodes and $L=8N$ directed links and periodic boundary conditions endowed with an Euclidean metric.
The  choice of an Euclidean metric is dictated by the assumption (already formulated in  Ref.\cite{bianconi2021topological}) that the Lorentzian nature of space-time might be fully accounted by the operators (such as the Dirac operators) acting on this space rather than being a intrinsic property of the discrete network describing space-time. This allows us to avoid the challenge to define space-like and time-like links. In our approach the only difference between space-like and time-like links will depend on the choice of the Dirac operator that acts on them.

As a consequence of this, throughout the paper we will adopt the Einstein convention of repeated indices with $a_{\mu}{b}^{\mu}$ indicating $\sum_{\mu=1}^4a_{\mu}b_{\mu}$.

As in classical dynamics the configuration space is formed by the positions and the velocities of the particles we assume that the topological wave function ${\bm  \Psi}$  is the direct sum between a  two $0$-cochains and a two $1$-cochains  defined on each directed link of the network. Let us indicate with $C_0$  the set of $0$-cochains and with  and $C_1$ the set of directed  $1$-cochains \cite{grady2010discrete}. With this notation we have $\bm\Psi\in C_0\oplus C_0\oplus C_1\oplus C_1$. Hence $\bm\Psi=\bm\chi\oplus \bm\psi$ or alternatively
\bea
{\bm \Psi}=\left(\begin{array}{c}\bm\chi\\\bm\psi\end{array}\right).
\eea
The first component  $\bm\chi$ is formed by two $0$-cochains $\bm\chi\in C_0\oplus C_0$ and the second component $\bm\psi$ is formed by two directed 1-cochains $\bm\psi\in C_1\oplus C_1$
\bea
\bm\chi=\left(\begin{array}{c}\bm \chi_{+}\\\bm \chi_{-}\end{array}\right),\quad
\bm\psi=\left(\begin{array}{c} {\bm \psi_{+}}\\\bm\psi_{-}\end{array}\right).\eea Therefore   $\bm \chi_{\pm}$ are $0$-cochains formed by elements defined on the $N$ nodes of the network, while  $\bm \psi_{\pm}$ are directed 1-cochains 
 formed by elements defined on the links $L$ with any given direction
\bea\bm\chi_{\pm}=\left(\begin{array}{c}\chi_{1,\pm},\\\chi_{2\pm},\\\vdots\\ \chi_{N,\pm}\end{array}\right),\quad 
\bm\phi_{\pm}=\left(\begin{array}{c}\phi_{\ell_1,\pm}\\ \psi_{\ell_2,\pm}\\\vdots \\\psi_{\ell_{L},\pm}\end{array}\right).
\eea
Here we distinguish between $\psi_{[i,j],\pm}$ and $\psi_{[j,i],\pm}$, in order to  interpret  $\psi_{{[i,j]},\pm}$ as a ``flux going from $i$ to $j$"  thus indicating a  variable localized on the node $i$ while 
 $\psi_{{[j,i]},\pm}$ indicates a "flux going from $j$ to $i$", which  is  instead a variable localized on node $j$. Note that in our directed setting  values of $\psi_{[i,j]\pm}$ with $\psi_{[i,j]\pm}\neq-\psi_{[j,i]\pm}$ are allowed. 
 
 This framework  can  treat a {\em local} field theory in which the  topological spinor  is defined on both nodes and links of the network. The part of the topological spinor localized on the node $i$ is given by  
 \bea
 \bm {\hat{ \Psi}}_i=\left(\begin{array}{c}\bm\chi_i\\\bm\psi_i\end{array}\right).
 \eea
where $\bm \chi_i$ is  formed by the elements of $\bm\chi$ localized on node $i$ and $\bm\psi_{i}=\left(\ldots {\bm\psi}_{[i,j]}\ldots\right)^{\top}$ is formed by   the direct sum of vectors ${\bm\psi}_{[i,j]}=(\psi_{[i,j],+},\psi_{[i,j],-})^{\top}$   indicating the fluxes from node $i$ to a generic  neighbour $j$.
Strictly speaking $\bm \psi_{i}$ is   a field associated to the fiber bundle defined at node $i$ (see for instance discussion of "vertex spaces" in Ref.\cite{post2009first}).

The scalar product between topological spinors is taken to be the $L^2$ norm
\bea
\Avg{\hat{\bm\Psi},\bm \Psi}= \hat{\bm\Psi}^{\dag}\bm \Psi.
\eea
Similarly the scalar product between the $0$-cochains and the scalar product between $1$-cochain is taken to be the $L^2$ norm
\bea
\Avg{\hat{\bm \chi},\bm \chi}= \hat{\bm \chi}^{\dag}\bm \chi, \quad \Avg{\hat{\bm\psi},\bm \psi}= \hat{\bm \psi}^{\dag}\bm \psi.
\label{scalar_01}
\eea
In  a ($3+1$)-dimensional lattice  we classify links in 4 classes: $x$-type, $y$-type, $z$-type and $t$-type links.
We consider the  $ L\times N$ weighted coboundary matrix $\bar{\bf B}_{\mu}$ of  type $\mu\in \{t,x,y,z\}$ links 
which is given by 
\bea
\bar{\bf B}_{\mu}={\bf G}_{[1]}^{-1/2}\bar{\bf B}_{\mu}^{(U)}{\bf G}_{[0]}^{1/2}
\eea
where $\bar{\bf B}_{\mu}^{(U)}$ is the $L\times N$ unweighted coboundary matrix  of elements
\bea
[{\bar{B}_{\mu}^{(U)}}]_{\ell i}=\left\{\begin{array}{cclcl}-1 &\mbox{if}& \ell=[i,j] & \& &\ell\  \mbox{is a  }\mu-\mbox{link} \\1&\mbox{if}& \ell=[j,i] & \& & \ell\  \mbox{is a  } \mu-\mbox{link},\end{array}\right.
\eea
and where ${\bf G}_{[0]}$ and ${\bf G}_{[1]}$ are the $N\times N$ and $L\times L$ metric matrices defined among  nodes and among links respectively.
The metric matrices ${\bf G}_{[0]}$ and ${\bf G}_{[1]}$ describe the geometry of the network and will be interpreted as metric degree of  freedom which are interacting with the matter field $\bm\Psi$.
hile in topology and exterior calculus \cite{desbrun2005discrete} typically these metric matrices are taken to be diagonal, with the diagonal elements of ${\bf G}_{[1]}^{-1}$ indicating the weights of the links and the diagonal elements of ${\bf G}_{[0]}^{-1}$ indicating the weights associated to the nodes, here we allow the metric matrices to be non diagonal and we allow their matrix elements to be complex.
In particular here and in the following we will take 
\bea
{\bf G}_{[0]}=e^{-{\bf A}^{(0)}},\quad {\bf G}_{[1]}=e^{-{\bf A}^{(1)}},
\eea
where the $N\times N$ matrix ${\bf A}^{(0)}$ and the $L\times L$ matrix  ${\bf A}^{(1)}$ describe the metric fields of our gauge theory. If the metric is flat then both ${\bf A}^{(0)}$ and ${\bf A}^{(1)}$ are proportional to the identity matrices of dimension $N\times N$ and $L\times L$ respectively. 

The matrices ${\bar{\bf B}}^{\star}_{\mu}$ are the Hodge-star of the  co-boundary matrices ${\bar{\bf B}}_{\mu}$ with respect to the standard $L^2$ norm and they are given by  \cite{grady2010discrete,baccini2022weighted,desbrun2005discrete,lim2020hodge}
\bea
{\bf \bar{B}}^{\star}_{\mu}={\bf \bar{B}}^{\dag}_{\mu}.
\eea
The weighted graph Laplacian operator \cite{chung1997spectral} in the  direction $\mu$ \cite{grady2010discrete,desbrun2005discrete,lim2020hodge,bianconi2021higher} describing diffusion from  nodes to nodes through links of type $\mu$ is given by  ${\bf L}_{\mu}={\bf \bar{B}}^{\star}_{\mu}{\bf \bar{B}}_{\mu}$.

In order to define the  Dirac operators let  us  first introduce the tensor product to the  Pauli matrices with the matrix ${\bf F}$,  $\bm\sigma_{\mu}\otimes {\bf F}$ with $\mu\in \{t,x,y,z\}$  as the matrices having  block structure given by 
\bea
\hspace*{-8mm}&&\bm\sigma_t \otimes {\bf F}=\bm\sigma_0\otimes{\bf F}=\left(\begin{array}{cc}{\bf F}&{\bf 0} \\ {\bf 0}&{\bf F}\end{array}\right),\ \ \bm\sigma_x\otimes {\bf F}=\left(\begin{array}{cc}{\bf 0}&{\bf F} \\ {\bf F}&0\end{array}\right),\ \ \nonumber \\ &&\bm\sigma_y\otimes {\bf F}=\left(\begin{array}{cc}{\bf 0}&-\mathrm{i}{\bf F} \\ \mathrm{i}{\bf F} &{\bf 0}\end{array}\right),\ \ \bm\sigma_z \otimes {\bf F}=\left(\begin{array}{cc}{\bf F}& {\bf 0}\\{\bf 0}& -{\bf F}\end{array}\right).
\label{Pauli}
\eea
The exterior derivative $d_{\mu}$ in the direction $\mu$ and its adjoint $d^{\star}_{\mu}$ are given by 
\bea
d_{\mu}=\left(\begin{array}{cc}
0&0\\
\bm\sigma_0\otimes {\bf \bar{B}}_{\mu}&0
\end{array}\right),\quad 
d^{\star}_{\mu}=\left(\begin{array}{cc}0&\bm \sigma_0\otimes{\bf \bar{B}}_{\mu}^{\star}\\
0&0
\end{array}\right),
\label{ext_mu} 
\eea
for $\mu\in \{t,x,y,z\}$ leading to the directional Hodge-Dirac operator (see Appendix A for further details)
\bea
d_{\mu}+d^{\star}_{\mu}=\left(\begin{array}{cc}
0&\bm \sigma_0\otimes{\bf \bar{B}}_{\mu}^{\star}\\
\bm\sigma_0\otimes {\bf \bar{B}}_{\mu}&0
\end{array}\right).
\label{Hodge_Dirac}
\eea
We define the  directional Dirac operators $\partial_{\mu}$   as 
\bea
\partial_{\mu}=\bm\gamma_{\mu}(d_{\mu}^{\star}+d_{\mu}),
\eea
(note that here the indices are not contracted),
where the matrices $\bm\gamma_t$ and $\bm\gamma_{\mu}$ with $\mu\in \{x,y,z\}$ are given by 
\bea
\hspace{-20mm}\bm\gamma_{t}=\bm\gamma_{0}= \left(\begin{array}{cc}\bm \sigma_0\otimes{\bf I}_{N}&0\\
0& -\bm \sigma_0\otimes {\bf I}_{L}
\end{array}\right),\quad
\bm\gamma_{\mu}=-\iu \left(\begin{array}{cc}\bm \sigma_\mu\otimes {\bf I}_{N}&0\\
0& -\bm \sigma_\mu\otimes {\bf I}_{L}
\end{array}\right),
\label{g_mu}
\eea
with ${\bf I}_X$ indicating the identity matrix of dimension $X\times X$. 
We observe that given the definition of $d_{\mu}$ and $d_{\mu}^{\star}$ given by Eq.(\ref{ext_mu})  and the definition of the matrices $\bm\gamma_{\mu}$ given by Eq.(\ref{g_mu})  we obtain the anticommutator relations 
\bea
\{\bm\gamma_{\mu},(d_{\mu}+d_{\mu}^{\star})\}=0,
\label{anti_g_d}
\eea
valid for $\mu\in \{t,x,y,z\}$.
Finally we use  $\dirac$ to indicate
\bea
\dirac=\gamma^{\mu}(d_{\mu}+d^{\star}_{\mu}),
\eea
or alternatively $\dirac=\sum_{\mu}\partial_{\mu}$.
We observe that  the  anticommutator relation $\{\bm\gamma_{0},(d_{\mu}+d_{\mu}^{\star})\}=0,$ is  related to the supersymmetric interpretation \cite{witten1982supersymmetry} of the topological spinor where $0$-cochains are interpreted as bosonic components while $1$-cochains are interpreted as fermionic components and $\bm\gamma_0=(-1)^F$ \cite{post2009first,miranda2023continuum} with the Hodge-Dirac operator  mapping fermionic components to bosonic components and viceversa. Therefore the Hodge-Dirac operator is often interpreted as a supersymmetry operator. Note however that typically it is required  \cite{witten1982supersymmetry} that  the supersymmetry operators  anticommute with each other,  condition that in our setting does not hold as we will detail in the following.\\

From the above definition it follows that $\partial_t$ and  $\partial_{\mu}$ for $\mu\in \{x,y,z\}$  have  the block structure 
\bea
\hspace{-10mm}\partial_t=\left(\begin{array}{cc}0&\bm\sigma_{t}\otimes{\bf \bar{B}}^{\star}_t\\
-\bm\sigma_{t}\otimes{\bf \bar{B}}_t&0\end{array}\right),\quad
\partial_\mu=\left(\begin{array}{cc}0&-\textrm{i}\bm\sigma_{\mu}\otimes{\bf \bar{B}}^{\star}_{\mu}\\
\textrm{i}\bm\sigma_{\mu}\otimes{\bf \bar{B}}_{\mu}&0\end{array}\right).
\eea 
%Here $\bar\mathcal{B}_{\mu}$ and $\bar\mathcal{B}_\mu$ are given by 
%\bea
%\bar\mathcal{B}_{t}=\bm\sigma_{t}({\bf \bar{B}}_t),\ \ \bar\mathcal{B}_{x}=\bm\sigma_{x}({\bf \bar{B}}_x),\ \ 
%\bar\mathcal{B}_{y}=\bm\sigma_{y}({\bf \bar{B}}_y),\ \ \bar\mathcal{B}_{z}=\bm\sigma_{z}({\bf \bar{B}}_z),\nonumber \\
%\bar\mathcal{B}^{\star}_{t}=\bm\sigma_{t}({\bf \bar{B}}^{\star}_t),\ \ \bar\mathcal{B}^{\star}_{x}=\bm\sigma_{x}({\bf \bar{B}}^{\star}_x),\ \ 
%\bar\mathcal{B}^{\star}_{y}=\bm\sigma_{y}({\bf \bar{B}}^{\star}_y),\ \ \bar\mathcal{B}^{\star}_{z}=\bm\sigma_{z}({\bf \bar{B}}^{\star}_z).
%\eea
This definition is such that $\textrm{i}\bm\gamma_0\partial_{\mu}$ is Hermitian for $\mu\in \{x,y,z\}$ and anti-Hermitian for $\mu=t$.
In particular we obtain 
\bea
\hspace{-10mm}\textrm{i}\bm\gamma_0\partial_\mu=\left(\begin{array}{cc}0&\bm\sigma_{\mu}\otimes{\bf \bar{B}}^{\star}_{\mu}\\
\bm\sigma_{\mu}\otimes{\bf \bar{B}}_{\mu}&0\end{array}\right),\quad 
\textrm{i}\bm\gamma_0\partial_t=\left(\begin{array}{cc}0&\textrm{i}\bm\sigma_{t}\otimes {\bf \bar{B}}^{\star}_t\\
\textrm{i}\bm\sigma_{t}\otimes{\bf \bar{B}}_t&0\end{array}\right).
\eea  
The Dirac action  $\mathcal{S}_D$ is defined as 
\bea
\mathcal{S}_{D}=\bar{\bm \Psi}(\textrm{i}\dirac-m){\bm\Psi}.
\label{L_dirac}
\eea
where $\bar{\bm\Psi}=\bm\Psi^{\dag}\bm\gamma_0$. This notation implies contraction of all the indices in the topological spinor, indicating its values on all the nodes and all the directed links of the network. Moreover we observe that in this work we do not  consider the second quantization of the fields, so  the Dirac action $\mathcal{S}_D$ is for us a $c$-number. 
The dynamical equations associated to this action is  the Dirac equation  and its adjoint. The Dirac equation is obtained by deriving the action with respect to each element $\bar{\Psi}_r$ of the topological spinor $\bar{\bm\Psi}$ getting
\bea
(\textrm{i}\dirac-m) \bm{\Psi}=0,
\label{dirac}
\eea 
and its adjoint is obtained by deriving the action with respect to each element $\Psi_r$ of the topological spinor $\bm\Psi$ is given 
\bea
(-\textrm{i}\dirac-m) \bm{\bar\Psi}=0.
\label{dirac_adjoint}
\eea
The spectrum of the Dirac operator is 
\bea
\mathcal{E}^2=m^2 
\eea
where $\mathcal{E}^2$ is the eigenvalue of the D'Alabertian $\square={\bf L}_t-\sum_{\mu\in \{x,y,z\}}{\bf L}_{\mu}$.
If $[{\bf L}_{t},{\bf L}_{x}+{\bf L}_y+{\bf L}_z]=0$ we recover the relativistic dispersion 
\bea
E^2=m^2+{p}^2
\eea
where  $E^2$ is the eigenvalue of the operator ${\bf L}_t$ and $p^2$ is the eigenvalue of ${\bf L}_{x}+{\bf L}_y+{\bf L}_z$. In the case in which   $[{\bf L}_\mu,{\bf L}_\nu]=0$  for any $\mu\in \{t, x,y,z\}$ and $\nu\in \{t,x,y,z\}$ we have
\bea
E^2=m^2+|{\bf p}|^2
\eea
where ${\bf p}=(p_x,p_y,p_z)$ with $p_\mu$ indicating  the eigenvalue of the directional Dirac operator  $ {\partial}_{\mu}$ with $\mu\in \{x,y,z\}$.
\section{Non-commutative Dirac operators}
\label{Sec:comm}
The square of the spatial Dirac operators $\partial_{\mu}$ with $\mu\in \{x,y,z\}$ are given by the Laplacian matrices $\mathcal{L}_{\mu}$
\bea
\partial_{\mu} \partial_{\mu}={\mathcal L}_{\mu}=\left(\begin{array}{cc}\bm\sigma_0\otimes({\bar{\bf B}}_{\mu}^{\star}{\bar{\bf B}_{\mu}})&0\\0&\bm\sigma_0\otimes ({\bar{\bf B}}_{\mu}{\bar{\bf B}}_{\mu}^{\star})\end{array}\right)
\eea
while for the square of the temporal Dirac operator is given by $-\mathcal{L}_t$, i.e.
\bea
\partial_{t} \partial_{t}=-{\mathcal L}_{t}=-\left(\begin{array}{cc}\bm\sigma_0\otimes({\bar{\bf B}}_{t}^{\star}{\bar{\bf B}_{t}})&0\\0&\bm\sigma_0\otimes({\bar{\bf B}}_{t}{\bar{\bf B}}_{t}^{\star})\end{array}\right).
\eea
The anticommutators and the commutators of the  Dirac operators are  non zero in general, even for a  flat metric. In particular for $\mu\in \{x,y,z\}$, $\nu\in \{x,y,z\}$ with $\mu\neq \nu$
\bea
\{\partial_{\mu}, \partial_{\nu}\}= -\left(\begin{array}{cc}0&0\\0&\epsilon_{\mu\nu\theta}\left[\bm\sigma^{\theta}\otimes [{\bf B}^{(M)}]^{\theta}\right]\end{array}\right).
\eea
where the magnetic field ${\bf B}_{\theta}^{\bf (M)}$ is given by the $L\times L$ matrix given by 
\bea
{\bf B}_x^{(M)}=-\iu(\bar{\bf B}_{y}\bar{\bf B}_{z}^{\star}-\bar{\bf B}_{z}\bar{\bf B}_{x}^{\star}),\nonumber \\
{\bf B}_y^{(M)}=-\iu(\bar{\bf B}_{z}\bar{\bf B}_{x}^{\star}-\bar{\bf B}_{x}\bar{\bf B}_{z}^{\star}),\nonumber \\
{\bf B}_z^{(M)}=-\iu(\bar{\bf B}_{x}\bar{\bf B}_{y}^{\star}-\bar{\bf B}_{y}\bar{\bf B}_{x}^{\star}).
\eea
It follows that the magnetic field ${\bf B}^{(M)}_{\theta}$ is non vanishing in general, instead is given by anti-symmetric matrix.

For $\mu=t$ and $\mu\in \{x,y,z\}$ we have instead
\bea
\{\partial_{t},\partial_{\mu}\}=
\left(\begin{array}{cc}0&0\\0&\iu\bm\sigma_\mu\otimes (\bar{\bf B}_{t}{\bar{\bf B}}_{\mu}^{\star}+{\bar{\bf B}}_{\mu}{\bar{\bf B}}_{t}^{\star}).\end{array}\right)
\eea
The commutator of the  Dirac operators in the spatial directions $\mu,\nu\in \{x,y,z\}$ give 
\bea
[\partial_{\mu},\partial_{\nu}]=-\iu
\left(\begin{array}{cc}0&0\\0& \epsilon_{\mu\nu\theta}\left[\bm\sigma^{\theta}\otimes (\bar{\bf B}_{\mu}{\bar{\bf B}}_{\nu}^{\star}+{\bar{\bf B}}_{\nu}{\bar{\bf B}}_{\mu}^{\star})\right].\end{array}\right).
\label{c1}
\eea
The commutator of the Dirac operators in the temporal direction $t$ and in the spatial directions $\mu\in \{x,y,z\}$ gives 
\bea
[\partial_{t},\partial_{\mu}]=\iu 
\left(\begin{array}{cc}0&0\\0&\bm\sigma_\mu\otimes (\bar{\bf B}_{t}{\bar{\bf B}}_{\mu}^{\star}-{\bar{\bf B}}_{\mu}{\bar{\bf B}}_{t}^{\star})\end{array}\right).
\label{c2}
\eea
The commutators $[\partial_{\mu},\partial_{\nu}]$ with $\mu,\nu\in \{t,x,y,z\}$ will be interpreted as the curvature tensor of our theory and their contraction will be used in Sec. $\ref{Sec:metric}$ to formulate the action for the metric degrees of freedom.
While the commutators of the Dirac operators do not vanish, we have however
\bea
\partial_\rho[\partial_\mu,\partial_\nu]=0,
\eea
for $\rho\neq \mu,\rho\neq\nu$.

\section{Non-relativistic limit of the Dirac equation}

In this section our goal is to carry out the non-relativistic limit \cite{ryder1996quantum} for the Dirac equation for $m>0$. The main difference with the textbook calculation is that the  topological spinor has now a geometrical interpretation. As we will see this difference will carry notable consequences for the non-relativist limit of our equation. What we will see is that in the non-relativistic limit the wave function defined on the spatial section of the fiber bundle obeys the Schr\"odinger equation while the wave function defined on the nodes obeys the Klein-Gordon equation and is not negligible.  
In order to discuss the non-relativistic limit of our Dirac equation, let us  consider the case in which ${\bf L}_t$ commutes with ${\bf L}_{x}+{\bf L}_{y}+{\bf L}_z$ but the spatial graph Laplacians do not commute with each other, due to the non-trivial metric matrices, i.e. $[{\bf L}_{\mu},{\bf L}_{\nu}]\neq 0$ for $\mu,\nu\in \{x,y,z\}$. We consider therefore a Dirac wave-function associated with the eigenvalues $E$  of the operator $\iu\partial_t$ with $E\sim m$ and we characterize the eigenvectors of the Dirac equation.
The Dirac equation $(\ref{dirac})$ can be expressed as and equation for $\bm\chi$ and $\bm\psi$ with $\bm\Psi=(\bm\chi,\bm\psi)^{\top}$ as
is given by 
\bea
&&\sum_{\mu\{x,y,z\}}(\bm\sigma_{\mu}\otimes \bar{\bf B}_{\mu}^{\star})\bm\psi+\iu (\bm\sigma_0\otimes \bar{\bf B}_{t}^{\star})\bm\psi-m\bm\chi=0,\label{d_chi}\\
&&-\iu(\bm\sigma_{0}\otimes \bar{\bf B}_{t})\bm\chi-m\bm\psi_t=0,\label{d_psi_t}\\
&&-(\bm\sigma_{\mu}\otimes\bar{\bf B}_{\mu})\bm\chi-m\bm\psi_{\mu}=0,\label{d_psi_mu} 
\eea
where $\bm\psi_\mu$ indicates the vector constructed from $\bm\psi$ by retaining only the elements defined on links of direction $\mu$ for any possible choice of $\mu\in\{t,x,y,z\}$. 
Following an argument similar to the one provided in \cite{bianconi2021topological}, i.e. substituting Eq.(\ref{d_psi_mu}) and Eq.(\ref{d_psi_t}) into Eq.(\ref{d_chi}) it is easy to show that $\bm\chi$ satisfies the Klein-Gordon equation
\bea
(\bm\sigma_{0}\otimes \square)\bm\chi=m^2\bm\chi.
\eea
where $\square={\bf L}_t-\sum_{\mu\in \{x,y,z\}}{\bf L}_{\mu}$
Therefore $\bm\chi$ is an eigenvector of $\bm\sigma_0\otimes \square$. Since the d'Alabertian commutes with ${\bf L}_t$, $\bm\chi$ can also be chosen to be an eigenvector of $\bm\sigma_0\otimes {\bf L}_t$ with eigenvalue $E^2$.
If we now consider Eq.(\ref{d_psi_t}) we obtain
\bea
\bm\psi_{t}=-\frac{1}{m}\iu(\bm\sigma_{0}\otimes\bar{\bf B}_{t})\bm\chi
\eea
%\end{document}
Hence, using this relation we obtain
\bea
\iu (\bm\sigma_0\otimes\bar{\bf B}_{t}^{\star})\bm\psi=\frac{1}{m}(\bm\sigma_0\otimes {\bf L}_t)\bm\chi=\frac{E^2}{m}\bm\chi.
\eea
Inserting this equation in Eq.(\ref{d_chi}) we obtain 
\bea
-\sum_{\mu\in \{x,y,z\}}(\bm\sigma_{\mu}\otimes \bar{\bf B}_{\mu}^{\star})\bm\psi=\frac{E^2-m^2}{m}\bm\chi,
\label{nodes_bundle}
\eea
This is the relation between the component $\bm\chi$ and $\bm\psi$ and for $E\sim m$, $\bm \chi=O(1/(E-m))$
For $E\sim m$ we have $E^2-m^2\simeq 2m(E-m)$ and hence
\bea
\bm\chi\simeq -\frac{1}{2(E-m)}\sum_{\mu\in \{x,y,z\}}(\bm\sigma_{\mu}\otimes\bar{\bf B}_{\mu}^{\star})\bm\psi.
\eea
In follows that inserting this expression into the Eq.(\ref{d_psi_mu}) we obtain that the spatial sector on the wave function on the fiber bundle $\bm\psi_s=\bm\psi_x+\bm\psi_y+\bm\psi_z$ obeys
\bea
(E-m)\bm\psi_s=\frac{1}{2m}\sum_{\mu,\nu\in \{x,y,z\}}(\bm\sigma_{\mu}\otimes \bar{\bf B}_{\mu})(\bm\sigma_{\nu}\otimes\bar{\bf B}_{\nu}^{\star})\bm\psi_s.
\eea
Therefore we obtain that $\bm \psi$ follows the Schr\"odinger equation 
\bea
(E-m)\bm\psi_s=\frac{1}{2m}\left[\sum_{\mu}(\bm\sigma_0\otimes {\bf L}_\mu)-\sum_{\theta}(\bm\sigma_{\theta}\otimes {\bf B}_\theta^{\bf (M)})\right]\bm\psi_s
\eea
Therefore the  in the non-relativistic limit the component on the fiber bundle follows Schr\"odinger equation with giromagnetic constant $2$ while the component on the nodes obeys the Klein-Gordon equation. The node component $\bm\chi$ and the component on the fiber bundle $\bm\psi$ are related by Eq.(\ref{nodes_bundle}). Therefore $\bm\chi$ is not negligible with respect to $\bm\psi$. 
We believe that these results might be related to the supersymmetric interpretation of the Dirac operator  and might guide  the most suitable second quantization of this theory.
%\end{document}
\section{Weyl equation for topological spinors}
The  Weyl equation for topological spinors is obtained when the mass is zero, i.e. $m=0$.
In this case the topological Weyl equation reads 
\bea
&&\sum_{\mu\in \{x,y,z\}}(\boldsymbol{\sigma}_{\mu}\otimes \bar{\bf B}_{\mu}^{\star})\bm\psi+\iu (\bm{\sigma}_0\otimes \bar{\bf B}_t^{\star})\bm\psi=0 \label{weyl_psi}\nonumber \\
&&-(\bm\sigma_{\mu}\otimes \bar{\bf B}_{\mu})\bm\chi=0,\label{weyl_chi1} \\
&&-\iu(\bm\sigma_{0}\otimes \bar{\bf B}_{t})\bm\chi=0.\label{weyl_chi2}
\eea 
%\end{document}
This equation implies that $\bm\psi$ satisfies Eq.(\ref{weyl_psi}) and that $\bm\chi$ belongs to the intersection of the kernels of  $\bm\sigma_{\mu}\otimes \bar{\bf B}_{\mu}$, i.e.
\bea
\bm\chi\in \cap_{\mu\in \{t,x,y,z\}}\ \mbox{ker}(\bm\sigma_{\mu}\otimes \bar{\bf B}_{\mu})
\eea
\section{Global transformations }
We observe that the Dirac action $\mathcal{S}_{D}$ is invariant under a global phase transformation
\bea
\bm\Psi\to  e^{-\iu e \Lambda{\bf I}_{M}}  \bm\Psi,
\eea
where $\Lambda$ is taken to be a arbitrary  real constant and $M=2N+2L$.
In the first order approximation for $\Lambda\ll 1$ we have  $\bm \Psi\to \bm \Psi+\delta \bm \Psi$ with  
\bea
\delta \bm \Psi=-\iu e \Lambda \bm\Psi
\eea
We have that the Dirac action $\mathcal{S}_D$ is invariant under this transformation, i.e.
\bea
\delta \mathcal{S}_D=-\iu e\bar{\bm \Psi}(\iu\dirac-m) \Lambda\bm \Psi+\iu e\bar{\bm \Psi}\Lambda(\iu\dirac-m)\bm\Psi=0
\eea
Now $\dirac=\bm \gamma^{\mu}(d_{\mu}+d_{\mu}^{\star})$ with $d_{\mu}+d_{\mu}^{\star}$ obeying the anti-commutator relations with the $\gamma_{\mu}$ matrix given by  Eq. (\ref{anti_g_d}). Therefore we  have 
\bea
\delta \mathcal{S}_D=-e\Lambda\left[\Avg{\dirac \bar{\bm \Psi}^{\dag}, \bm\Psi}+\Avg{\bm  \bar{\bm \Psi}^{\dag}, \dirac\bm\Psi}\right]=0,
\eea
where the scalar product $\Avg{\cdot,\cdot}$ between two topological spinors is taken to be the standard $L^2$ norm.
Hence we obtain the following relation 
\bea
\Avg{\dirac \bar{\bm \Psi}^{\dag}, \bm\Psi}+\Avg{\bm  \bar{\bm \Psi}^{\dag} , \dirac\bm\Psi}=0.
\eea
%generalizing in our context the notion of conserved current.
In the next section we will consider gauge transformations which make the Dirac action also invariant under local transformation of the spinor.

\section{Gauge transformations}
In the this paragraph we will construct a topological (Abelian) transformation in order to guarantee  that the action $\mathcal{S}_D$ is invariant under the local $U(1)$ symmetry
\bea
\bm \Psi\to  e^{-\iu e\bm \Lambda}  \bm \Psi.
\eea
where $\bm\Lambda$ is an arbitrary  diagonal $M\times M$ matrix with $M=2N+2L$.

We will also consider non-Abelian transformations acting on 
each local spinor ${\bm\psi}_{i,\pm}$. Let us indicate with ${\bf \hat{J}}_{i}^{\alpha \beta}$ the generator of the Lorentz group $SL(2,C)$ group acting on the axis $({\alpha, \beta})$ at node $i$ which can be interpreted as a transformation of  the fiber bundle defined at node $i$. 
We will consider the set of  transformations in which the local spinors obey 
\bea
\bm{\psi}_{i\pm}\to e^{-\iu g {\bf \hat{J}}_i^{{\alpha \beta}}\Theta_{i\alpha\beta}} \bm{\psi}_{i\pm}
\eea
where for each value of $i$, $\Theta_{{i\alpha \beta}}$ are a set of $6$ $0$-cochains associated to the node $i$ of the network.
In this paragraph we will derive the  gauge transformations on the metric matrices ${\bf G}_{[0]}$ and ${\bf G}_{[1]}$ that will ensure the invariance of the action $\mathcal{S}_D$ under these transformations.

\subsection{Abelian Transformations}

Let us study the action $\mathcal{S}_D$ under the Abelian transformation 
\bea
\bm\Psi\to e^{-ie\bm\Lambda}\bm\Psi,
\eea
where we assume that $\bm\Lambda$ is an arbitrary diagonal $M\times M$  matrix with  block structure
\bea
e^{-\iu e\bm\Lambda}=\left(\begin{array}{cccc}e^{-\iu e\bm \Lambda_N}&0&0&0\\
0&e^{-\iu e\bm \Lambda_N}&0&0\\
0&0&e^{-\iu e\bm \Lambda_L}&0\\0&0&0&e^{-\iu e\bm \Lambda_L}\end{array}\right)
\eea
where $\bm \Lambda_N$ and $\bm\Lambda_L$ are  diagonal matrices of size $N\times N$ and $L\times L$ respectively.

By considering the action $\mathcal{S}_D$ defined in Eq.(\ref{L_dirac})
we observe that the action is clearly invariant under the Abelian transformation, provided the Dirac operator transform according to 
\bea
\dirac\to e^{-\iu e\bm\Lambda}\dirac e^{\iu e\bm\Lambda},
\label{dirac_transf}
\eea
which for small $\bm\Lambda$ implies  the gauge transformation 
\bea
\dirac\to \dirac-\iu e\A,
\eea
with 
\bea
\A=[\bm \Lambda,\dirac].
\eea
This implies that the coboundary operator  and the metric matrices are  transformed as 
\bea
\bar{\bf B}_{\mu}&\to& e^{-ie\bm \Lambda_L}\bar{\bf B}_{\mu}e^{\iu e\bm\Lambda_N},\nonumber \\
{\bf G}_{[0]}&\to& {\bf G}_{[0]}e^{2\iu e\bm \Lambda_N},\nonumber \\
{\bf G}_{[1]}&\to &{\bf G}_{[1]}e^{2\iu e\bm \Lambda_L},
\eea
while  before and after the transformation we  have
\bea
\bar{\bf B}^{\star}_{\mu}=\bar{\bf B}_{\mu}^{\dag}.
\label{HS}
\eea
This invariance implies that the graph Laplacian matrices can be expressed as ${\bf L}_{\mu}={\bf\bar{B}}_{\mu}^{\star}{\bf \bar{B}}_{\mu}$ both before and after the transformation. Moreover the trace of the product of graph Laplacians (including also graph Laplacians of different direction) is invariant under these transformations.

\subsection{Non-Abelian transformations}
Let us study the action $\mathcal{S}_D$ under the non-Abelian transformation 
\bea
\bm\Psi\to {S}(\{\bm \Theta_{\alpha\beta}\})\bm\Psi,
\label{S}
\eea
where we assume that the $M\times M$ matrix ${S}(\{\bm \Theta_{\alpha\beta}\})$ has block structure
\bea
{S}(\{\bm\Theta_{\alpha\beta}\})=\left(\begin{array}{cccc}1&0&0&0\\
0&1&0&0\\
0&0&e^{-\iu g\bm J^{\alpha \beta}_i\Theta_{\alpha \beta}^i}&0\\0&0&0&e^{-\iu g\bm J^{\alpha \beta}_i\Theta_{\beta \alpha}^i}\end{array}\right)
\eea
where for a given pair of $\alpha,\beta\in \{t,x,y,z\}$ with $\alpha\neq \beta$, $\bm J_{i}^{\alpha\beta}$ are $L\times L$ matrices that act of $\bm\psi_{\pm}$. These matrices $\bm J_{i}^{\alpha\beta}$  are obtained  the $4\times 4$ generators ${ \hat{\bm J}}_i^{\alpha\beta}$ of the Lorentz transformations of $\bm\psi_{i\pm}$ by acting trivially on all the other components of  $\bm \psi_{\pm}$ that are not localized on node $i$, and  $\bm \Theta_{\alpha\beta}=(\Theta_{\alpha\beta}^1,\Theta_{\alpha\beta}^1,\ldots, \Theta_{\alpha\beta}^N)$ are determining the gauge with $\Theta^i_{\alpha\beta}\in \mathbb{R}$.
We observe that the action $\mathcal{S}_D$  is clearly invariant under the non-Abelian transformations, provided the Dirac operator transforms according to 
\bea
\dirac\to {S}(\{\bm\Theta_{\alpha\beta}\})\dirac {S}^{-1}(\{\bm\Theta_{\alpha\beta}\}),
\label{dirac_transf}
\eea
which for small $\bm\Theta$ implies  the gauge transformation 
\bea
\dirac\to \dirac-\iu g\B,
\eea
with 
\bea
\B=[\mathcal{J}^{\alpha \beta}_i\Theta_{\alpha \beta}^i,\dirac],
\eea
where $\mathcal{J}_i^{\alpha\beta}$ is a $M\times M$ matrix of block structure
\bea
\mathcal{J}_i^{\alpha\beta}=\left(\begin{array}{cccc}
1&0&0&0\\
0&1&0&0\\
0&0&\bm J_i^{\alpha\beta}&0\\
0&0&0&\bm J_i^{\alpha\beta}
\end{array}\right).
\eea
This implies that the coboundary operator and the metric matrices are  transformed as 
\bea
\bar{\bf B}_{\mu}&\to& e^{-\iu g\bm J^{\alpha \beta}_i\Theta_{\alpha \beta}^i}\bar{\bf B}_{\mu},\nonumber \\
{\bf G}_{[0]}&\to & {\bf G}_{[0]},\nonumber \\
{\bf G}_{[1]}&\to &{\bf G}_{[1]}e^{2\iu g\bm J^{\alpha \beta}_i\Theta_{\alpha \beta}^i},
\eea
while we have as before the condition that the transformation does not modify Eq.(\ref{HS}) and hence the graph Laplacians are always expressed as ${\bf L}_{\mu}={\bf\bar{B}}_{\mu}^{\star}{\bf \bar{B}}_{\mu}$. Note that the trace of product of graph Laplacians is invariant under these transformations.

\subsection{Combining Abelian and Non-Abelian transformations}
Let us now consider the following combined transformation 
\bea
\bm\Psi\to \mathcal{C}(\bm \Lambda,\{\bm \Theta_{\alpha\beta}\})\bm\Psi,
\eea
where using the same notation as in the previous two paragraphs, we assume that the $M\times M$ matrix $\mathcal{C}(\bm \Lambda,\{\bm \Theta_{\alpha\beta}\})$ has block structure
%\begin{widetext}
\bea
\hspace*{-25mm}\mathcal{C}(\bm \Lambda,\{\bm \Theta_{\alpha\beta}\})=\left(\begin{array}{cccc}ce^{-\iu e\bm \Lambda_N}&0&0&0\\
0&e^{-\iu e\bm \Lambda_N}&0&0\\
0&0&e^{-\iu(e\bm \Lambda_L+g\bm J^{\alpha \beta}_i\Theta_{\alpha \beta}^i)}&0\\0&0&0&e^{-\iu(e\bm \Lambda_L+g\bm J^{\alpha \beta}_i\Theta_{\beta \alpha}^i)}\end{array}\right).
\label{dirac_trans_c}
\eea
%\end{widetext}

We observe that the action  $\mathcal{S}_D$ is invariant under the  transformation given by Eq.(\ref{dirac_trans_c}), provided the Dirac operator transform according to 
\bea
\dirac\to \mathcal{C}(\bm \Lambda,\{\bm \Theta_{\alpha\beta}\})\ \dirac \ \mathcal{C}^{-1}(\bm \Lambda,\{\bm \Theta_{\alpha\beta}\}),
\label{dirac_transf3}
\eea
which for infinitesimal transformations implies  the gauge transformation 
\bea
\dirac\to \dirac-\iu e\A-\iu g\B,
\eea
with $\A$ and $\B$ defined above.

The invariance  of the action $\mathcal{S}_D$ under the gauge  Eq.(\ref{dirac_transf3}) implies that the coboundary operator  and the metric matrices are  transformed as 
\bea
\bar{\bf B}_{\mu}&\to& e^{-\iu e\bm \Lambda_L-\iu g\bm J^{\alpha \beta}_i\Theta_{\alpha \beta}^i}\bar{\bf B}_{\mu}e^{\iu e\bm \Lambda_N},\nonumber \\
{\bf G}_{[0]}&\to &{\bf G}_{[0]}e^{2\iu e\bm \Lambda_N},\nonumber \\
{\bf G}_{[1]}&\to &{\bf G}_{[1]}e^{\iu 2(e\bm \Lambda_L+g\bm J^{\alpha \beta}_i\Theta_{\alpha \beta}^i)}.
\eea
We observe however that in general $\mathcal{C}(\bm \Lambda,\{\bm \Theta_{\alpha\beta}\})\neq \mathcal{K}(\bm \Lambda,\{\bm \Theta_{\alpha\beta}\})=e^{i\bm\Lambda}\mathcal{S}(\{\bm\Theta_{\alpha\beta}\})$. Therefore ensuring invariance of the action under the transformation 
\bea
\bm\Psi\to \mathcal{C}(\bm \Lambda,\{\bm \Theta_{\alpha\beta}\})\bm\Psi
\eea
is not equivalent to ensuring invariance under the transformations
\bea
\bm\Psi\to \mathcal{K}(\bm \Lambda,\{\bm \Theta_{\alpha\beta}\})\bm\Psi=e^{i\bm\Lambda}\mathcal{S}(\{\bm\Theta_{\alpha\beta}\})\bm\Psi\eea
since $\bm\Lambda$ and $\mathcal{J}_i^{\alpha\beta}$ do not commute in general.
\section{Schwinger-Dyson equations}
In this section we aim at deriving the Schwinger-Dyson equations\cite{peskin2018introduction} for our  gauge theory.
Hence we consider first the case of Abelian and then the case of non-Abelian transformations.
\subsection{Abelian transformations}
Let us consider the local Abelian transformation
\bea
\bm\Psi\to e^{-\iu e\bm\Lambda} \bm\Psi.
\label{trs0}
\eea
In the first order approximations these transformations read
\bea
\bm\Psi\to\bm\Psi-\iu e\bm\Lambda \bm\Psi.
\label{trs}
\eea
Under these transformations the Dirac action $\mathcal{S}_D$ will change according to 
\bea
\mathcal{S}_D\to \mathcal{S}_D+\delta \mathcal{S}_D\eea
where $\delta\mathcal{S}_D$ is given by \bea
\delta\mathcal{S}_D&=&- e\left[\Avg{\dirac\bm{\bar \Psi}^{\dag},\bm\Lambda \bm\Psi}+\Avg{\bm\Lambda\bm{\bar \Psi}^{\dag},\dirac \bm\Psi}\right].\eea
Now, indicating with $s$ the generic simplex of the network (being either  a node of a link), and taking into consideration that the matrix $\bm \Lambda$ is diagonal with diagonal elements $[\bm\Lambda]_{ss}=\Lambda_s$, we obtain
\bea
\bm\Lambda \bm\Psi=\sum_s \Lambda_s \bm\Psi_s,
\eea
where $\bm\Psi_s$ is the spinor obtained by $\bm\Psi$ in which only the element $s$ is retained while all the other elements are zero.
Using this expression for $\bm\Lambda \bm\Psi$ we can decompose $\delta \mathcal{S}_D$ as 
\bea
\delta \mathcal{S}_D&=&- e \sum_{s}\Lambda_s\left[\Avg{\dirac\bm{\bar \Psi}^{\dag},\bm\Psi_s}+\Avg{\bm{\bar \Psi}^{\dag}_s,\dirac \bm\Psi}\right].\
\eea
Indicating with $r$ and $r'$ two generic simplices  of the network, and with $\Psi_r,\bar{\Psi}_{r}$ the $r$  element of the spinors $\bm\Psi$ and $\bar{\bm\Psi}$ respectively, we  observe that   the functional integral 
\bea
\int \mathcal{D}\bm{\bar\Psi}\mathcal{D}\bm{\Psi}\mathcal{D}{\bf A}^{(1)}\mathcal{D}{\bf A}^{(0)} e^{\iu\mathcal{S}_D}  \Psi_{r}\bar{\Psi}_{r'}^{\top}
\eea
is invariant under the transformation given by Eq.(\ref{trs}).
Therefore we have
%\begin{widetext} 
\bea
0=\int \mathcal{D}\bm{\bar\Psi}\mathcal{D}\bm{\Psi}\mathcal{D}{\bf A}^{(1)}\mathcal{D}{\bf A}^{(0)}e^{\iu\mathcal{S}_D} \left\{\iu\delta\mathcal{S}_D\Psi_{r}\bar{\Psi}_{r'}- \iu e\Lambda_{r}\Psi_{r}\bar{\Psi}_{r'}+\iu e\Psi_{r}\Lambda_{r'}\bar{\Psi}_{r'}\right\}.
\eea
%\end{widetext}
Since $\bm\Lambda$ is arbitrary, we have obtained the  Schwinger-Dyson type equations
%\begin{widetext}
\bea
\int \mathcal{D}\bm{\bar\Psi}\mathcal{D}\bm{\Psi}\mathcal{D}{\bf A}^{(1)}\mathcal{D}{\bf A}^{(0)}e^{\iu\mathcal{S}_D}\left[\Avg{\dirac\bm{\bar \Psi}^{\dag}, \bm\Psi_{s}}+\Avg{\bm{\bar \Psi}_s^{\dag},\dirac \bm\Psi}\right]\bar{\Psi}_{r'} \Psi_{r}\nonumber \\
=\int \mathcal{D}\bm{\bar\Psi}\mathcal{D}\bm{\Psi}\mathcal{D}{\bf A}^{(1)}\mathcal{D}{\bf A}^{(0)}e^{\iu\mathcal{S}_D}[-\delta_{s,r}\bar{\Psi}_{r'} \Psi_{r}+\delta_{s,r'}\bar{\Psi}_{r'} \Psi_{r}].
\eea

\subsection{Non-Abelian transformations}
Let us consider the local non-Abelian transformation
\bea
\bm\Psi\to e^{-\iu g{\mathcal{J}}_i^{\alpha\beta}\Theta^i_{\alpha\beta}} \bm\Psi.
\label{trs0}
\eea
In the first order approximations these transformations read
\bea
\bm\Psi\to\bm\Psi+\delta \bm\Psi,\eea
with 
\bea
\delta \bm\Psi=-\iu g{\mathcal{J}}_i^{\alpha\beta}\Theta^i_{\alpha\beta} \bm\Psi.
\label{trs2}
\eea
Under these transformations  the Dirac action $\mathcal{S}_D$ transform  according to 
\bea
\mathcal{S}_D\to \mathcal{S}_D+\delta \mathcal{S}_D\eea
where $\delta\mathcal{S}_D$ is given by \bea
\delta\mathcal{S}_D&=&-g \Theta^{i}_{\alpha\beta} \left[\Avg{\dirac\bm{\bar \Psi}^{\dag},\mathcal{J}_i^{\alpha\beta} \bm\Psi}+\Avg{\mathcal{J}_i^{\alpha\beta}\bm{\bar \Psi}^{\dag},\dirac \bm\Psi}\right].\eea

Indicating with $r=[j,k]$ and $r'=[j',k']$ two generic links of the network, and by $\bar{\Psi}_{r'}$ and $\Psi_r$ the corresponding elements of the spinors $\bar{\bm \Psi}$ and $\bm\Psi$  we  observe that   the functional integral 
\bea
\int \mathcal{D}\bm{\bar\Psi}\mathcal{D}\bm{\Psi}\mathcal{D}{\bf A}^{(1)}\mathcal{D}{\bf A}^{(0)} e^{\iu\mathcal{S}_D}  \bar{\Psi}_{r'} \Psi_{r}
\eea
is invariant under the transformation given by Eq.(\ref{trs2}), therefore we have  
\bea
\hspace*{-23mm}0=\int \mathcal{D}\bm{\bar\Psi}\mathcal{D}\bm{\Psi}\mathcal{D}{\bf A}^{(1)}\mathcal{D}{\bf A}^{(0)}e^{\iu\mathcal{S}_D} \left\{\iu\delta\mathcal{S}_D\bar{\Psi}_{r'} \Psi_{r}+ \delta\bar{\Psi}_{r'}  \Psi_{r}+\bar{\Psi}_{r'} \delta\Psi_{r}\right\}.
\eea
Since $\bm\Theta$ is arbitrary we obtain the Swinger-Dyson type equations 
\bea
\hspace*{-22mm}\int \mathcal{D}\bm{\bar\Psi}\mathcal{D}\bm{\Psi}\mathcal{D}{\bf A}^{(1)}\mathcal{D}{\bf A}^{(0)}e^{\iu\mathcal{S}_D} \left[\Avg{\dirac\bm{\bar \Psi}^{\dag},\mathcal{J}_i^{\alpha\beta} \bm\Psi}+\Avg{\mathcal{J}_i^{\alpha\beta}\bm{\bar \Psi}^{\dag},\dirac \bm\Psi}\right]\bar{\Psi}_{r'} \Psi_{r}=\nonumber \\
 \hspace*{-22mm}=\int \mathcal{D}\bm{\bar\Psi}\mathcal{D}\bm{\Psi}\mathcal{D}{\bf A}^{(1)}\mathcal{D}{\bf A}^{(0)}e^{\iu\mathcal{S}_D}[\delta_{ij'}[\bar{\bm\Psi}{\mathcal{J}}_{j'}^{\alpha\beta}]_{r'=[j',k]} \Psi_{r}-\delta_{ij} \bar{\Psi}_{r'}[{\mathcal{J}}_j^{\alpha\beta}\bm \Psi]_{r=[j,k]}].
\eea
\section{The action for the metric degree of freedom}
\label{Sec:metric}
\subsection{General framework}
We can interpret the  Dirac operators $\partial_{\mu}$ as a topological version of a partial derivative in the direction $\mu$ coupled with a local rotation of the spinor.
However there is a major difference between the  Dirac operator and a partial derivative: mainly the difference is that the  Dirac operators associated to a different directions do not commute or anticommute (see discussion in Sec. \ref{Sec:comm} and in Ref. \cite{bianconi2021topological}).
Here we define the curvature tensor $F_{\mu\nu}$ and the tensor $G_{\mu\nu}$ as 
\bea
[\partial_{\mu},\partial_{\nu}]=\iu F_{\mu\nu},\nonumber \\
\{\partial_{\mu},\partial_{\nu}\}=\iu G_{\mu\nu}.
\eea
These tensors are dependent crucially on the metric matrices ${\bf G}_{[1]}=\exp(-{\bf A}^{(1)})$ and ${\bf G}_{[0]}=\exp(-{\bf A}^{(0)})$, i.e. it is determined by the metric fields ${\bf A}^{(1)}$ and ${\bf A}^{(0)}$
where for the moment we consider these two as independent fields (but actually the geometry might impose that they are related).
The Bianchi identities  are automatically satisfied for the curvature tensor,
\bea
\partial_{\rho}F_{\mu\nu}+\partial_{\mu}F_{\nu\rho}+\partial_{\nu}F_{\rho\mu}=0.
\eea
Let us recall the matrix form of the commutators $[\partial_{\mu},\partial_{\nu}]$ given by Eq.(\ref{c1}) and Eq.(\ref{c2}). From this expression it follows  that the curvature tensor $F_{\mu\nu}=-\iu[\partial_{\mu},\partial_{\nu}]$ is a $(2N+2L)\times (2N+2L)$ matrix that is non-zero only on the link-link sector.
%Remember however that each link is assigned a unique type $\mu$, with $\mu\in \{x,y,z,t\}$. As shown in Eq.() $F_{\mu\nu}$ 
Therefore $F_{\mu\nu}$ is not only endowed with the two indices $\mu$ and $\nu$  but t has elements determined by other two additional indices whose only non trivial choices are the ``link indices" $\rho=[{\ell_{\mu}\pm}]$ and $\sigma=[\ell_{\nu}\pm]$, where $\ell_{\mu}$ and $\ell_{\nu}$ indicate the links in direction $\mu$ and $\nu$ respectively.  We can then construct an action by ``contracting" the indices $\mu,\nu$ with the indices $\rho,\sigma$, i.e. we can consider the {\em Ricci scalar } $R$ given by 
\bea
R=\sum_{\mu,\nu}\sum_{\ell_{\mu},\ell_{\nu}}\sum_{n_{\mu},n_{\nu}\in {\pm}}F_{\mu\nu}(\ell_{\mu},n_{\mu};\ell_{\nu},n_{\nu})
\eea
where we have indicated with $F_{\mu\nu}(\ell_{\mu},n_{\mu};\ell_{\nu},n_{\nu})$ the elements of the matrix $F_{\mu\nu}$ calculated on the ``links elements" $\rho=[\ell_{\mu},n_{\mu}]$  and elements $\sigma=[\ell_{\nu},n_{\nu}]$.
Therefore this Ricci scalar leads to the  action $\mathcal{S}_R$ given by 
\bea
\mathcal{S}_R=R.
\eea
However this action is not invariant under the gauge transformations considered in the previous section.
In order to restore invariance under the mentioned gauge transformations one option is to couple $F_{\mu\nu}$  with the matter fields obtaining 
\bea
\mathcal{S}_C^{\prime}=-\Gamma_C\sum_{\mu\nu}\bar{\bm \Psi}F_{\mu\nu}\bm \Psi,
\eea
where $\Gamma_R$ is a constant. 
A similar argument can be made for the tensor $G_{\mu\nu}$, which might suggest considering the first term which is allowed by the considered transformations, i.e. 
\bea
\mathcal{S}_A^{\prime}=\Gamma_A\sum_{\mu\nu}\bar{\bm \Psi}G_{\mu\nu}\bm \Psi.
\eea
These considerations lead to the action $\mathcal{S}_R^{\prime}=\mathcal{S}_C^{\prime}+\mathcal{S}_A^{\prime}$ which couple metric degrees of freedom with matter fields. 
 
Alternatively we can contract the indices of the curvature tensor $F_{\mu\nu}$ by considering the product $F_{\mu\nu}F^{\mu\nu}$ and performing the trace.
We can thus consider the action $\mathcal{S}_G$ associated the metric field given by  
\bea
\mathcal{S}_G= -\Gamma_G\frac{1}{2}\mbox{Tr} \left(F_{\mu\nu}F^{\mu\nu}\right)&=&-\Gamma_G\mbox{Tr}\left(\sum_{\mu\neq \nu}(-1)^{\delta_{\mu,t}+\delta_{\nu,t}}{\bf L}_{\mu}{\bf L}_{\nu}\right),
\label{SG}
\eea
where $\Gamma_G$ is a constant.
We observe that all the transformations considered in the previous section leave invariant  the action $\mathcal{S}_G$.
Additionally is also possible to consider the action $\mathcal{S}_G^{\prime}$ including higher-order terms in the curvature 
\bea
\mathcal{S}_{G}^{\prime}&=&\frac{\Gamma_G^{\prime}}{2} \mbox{Tr}\sum_{\mu,\nu} (F_{\mu\nu})^4
=\Gamma_G^{\prime}
\sum_{\mu\neq\nu}\mbox{Tr}\left({\bf L}_{\mu}{\bf L}_{\nu}{\bf L}_{\mu}{\bf L}_{\nu}\right),
\label{SGp}
\eea
where $\Gamma_G^{\prime}$ is a constant.
The action $\mathcal{S}_G^{\prime}$ is also invariant under all the gauge transformations considered in the previous sections and  describes the contribution to the action given by the metrics fields  around each of the squares of the lattice.
Possibly one could extend this construction to actions including even larger powers of the curvature  $F$. However on a lattice the action $\mathcal{S}_{G}^{\prime}$ constitute the action involving the minimum power of $F$ while  allowing to take into consideration contributions along cyclic self-avoiding walks. Therefore we consider the  action $\mathcal{S}=\mathcal{S}_D+\mathcal{S}_R^{\prime}+\mathcal{S}_G+\mathcal{S}_G^{\prime}$ given by  
\bea
\hspace*{-23mm}{\mathcal S}=\bar{\bm \Psi} (\textrm{i}\dirac-m) \bm\Psi+\sum_{\mu,\nu}\bar{\bm \Psi}[-\Gamma_C F_{\mu\nu}+\Gamma_A G_{\mu\nu}]{\bm \Psi}- \frac{1}{2}\sum_{\mu\neq \nu}\mbox{Tr} \left[\Gamma_G(F_{\mu\nu})^2-\Gamma_G^{\prime} (F_{\mu\nu})^4\right].
\eea
The equations of motion are  obtained by deriving $\mathcal{S}$ with the respect ${\bf A}^{(1)}$ and ${\bf A}^{(0)}$, i.e. expressing with $A_{\gamma}$ the generic element of either ${\bf A}^{(1)}$ or ${\bf A}^{(0)}$. Under the assumption that ${\bf A}^{(1)}$ or ${\bf A}^{(0)}$ are unconstrained,  the equation of motion read,
\bea
\frac{\partial ({\mathcal{S}_D+\mathcal{S}_R^{\prime})}}{\partial  A_{\gamma}}+\frac{\partial (\mathcal{S}_G+\mathcal{S}_G^{\prime})}{\partial  A_{\gamma}}=0.\label{motion}
\eea
Here the first term depends on both the geometric degree of freedom and the Dirac fields while the second term depends exclusively on the metric degree of freedom (the $A_{\gamma}$).
These equations, together with Eq.(\ref{dirac}) and the gauge transformations define the dynamics of this Dirac gauge theory of the network.
Note that Eqs.(\ref{motion}) might be modified to  take into account geometric constraints between the metric fields ${\bf A}^{(1)}$ or ${\bf A}^{(0)}$.

\subsection{Simple example}
The study of the consequences of the equation of motion given by Eq. $(\ref{motion})$  is beyond the scope of this paper, but in the this  section their explicit expression is derived in the simple setting in which the metric matrices are diagonal.
To this end  we assume that the  metric matrix ${\bf G}_{[1]}^{-1}$ is diagonal and real valued with the diagonal element associated to link $[i,j]$ independent on the orientation of the link, i.e. the matrix element $[i,j],[i,j]$ is the same as the matrix element $[j,i][j,i]$. If the link $[i,j]$ is in the direction $\mu$ indicate the matrix elements as 
\bea
{\bf G}_{[1]}^{-1}([i,j],[i,j])={\bf G}_{[1]}^{-1}([j,i],[j,i])=w_{[i,j]}^{\mu}=e^{A^{\mu}_{[i,j]}}.
\eea 
where here for improving the clarity of the notation 
with $A^{\mu}_{[i,j]}$ 
we  indicate  the diagonal element ${A}^{(1)}([i,j][i,j])$ where $[i,j]$ is a link in the direction $\mu$, i.e. we omit the label $(1)$ and explicitly indicate the direction $\mu$ instead.
Moreover we assume that ${\bf G}_{[0]}$ is diagonal  with diagonal element associated to node $i$ given by the real number \bea
{\bf G}_{[0]}([i][i])=\frac{1}{w^{(0)}_i}=e^{-A^{(0)}_i}.\eea
For simplicity we assume that  ${\bf G}_{[0]}$ is not constrained by ${\bf G}_{[1]}$ but it is expected that the geometry will constraint the choices of the possible matrices ${\bf G}_{[0]}$ and ${\bf G}_{[1]}$ inducing other constraints in the equations of motion (Eq. $\ref{motion}$). 
In this setting  ${\bf \sigma}_{\mu}\otimes {\bar{\bf B}}_{\mu}$  acts on the elements of $\bm \chi$ as  
\bea
[({\bf \sigma}_{\mu}\otimes {\bar{\bf B}}_{\mu}){\bm\chi}]_{[i,j]}=\sqrt{w_{[i,j]}^{\mu}}\left(\frac{\bm\sigma_{\mu}\bm\chi_j}{\sqrt{w_j^{(0)}}}-\frac{\bm\sigma_{\mu}\bm \chi_i}{\sqrt{w_i^{(0)}}}\right).
\eea  
 while its adjoint acts on the elements of $\bm \psi$ as
\bea
[({\bf \sigma}_{\mu}\otimes{\bar{\bf B}}_{\mu}^{\star}){\bm\psi}]_i=\frac{1}{\sqrt{w^{0}_i}}\sum_{j\in N_{\mu}(i)}\sqrt{w_{[i,j]}^{\mu}}\left(\bm\sigma_{\mu}{\bm \psi}_{[j,i]}-\bm\sigma_{\mu}{\bm \psi}_{[i,j]}\right), 
\eea
where $N_{\mu}(i)$ indicates the set of nodes $j$ connected to $i$ via a $\mu$-type link.
The graph Laplacian ${\bf L}_{\mu}$ has diagonal elements given by  
\bea
[{\bf L}_{\mu}]_{ii}=2\frac{w^{\mu}_{[i,j']}+w^{\mu}_{[j'',i]}}{w^{(0)}_i}
\label{Lii}
\eea
where if $\mu=x$ the node $j'$ has $x$-coordinate $x_j=x_i+1$ and all the other coordinates equal to the ones of node $i$. Similarly, node $j''$ has $x$-coordinate $x_j=x_i-1$ and all the other coordinates equal to the ones of node $i$. Here we take periodic boundary conditions on the lattice. The generalization of the above expression for general direction $\mu$ is straightforward.
The non-diagonal elements of ${\bf L}_{\mu}$ are  given instead  by 
\bea
[{\bf L}_{\mu}]_{ij}=-2\frac{w^{\mu}_{[i,j]}}{\sqrt{w^{(0)}_iw^{(0)}_j}}.
\label{Lij}
\eea
as long as node $j$ is a connected to node $i$ by links of type $\mu$.

In this setting we have that $\mathcal{S}_R=R=0$ due to the symmetric choice of ${\bf G}_{[1]}^{-1}([i,j],[i,j])={\bf G}_{[1]}^{-1}([j,i],[j,i])$ however $\mathcal{S}_G$ and $\mathcal{S}_G^{\prime}$ will be non-zero and determined by the graph Laplacians ${\bf L}_{\mu}$.
We are interested now in deriving the explicit expression for the terms present in the equation of motion (Eq.(\ref{motion})). In the absence of matter fields, i.e. when $\bm\Psi={\bf 0}$, the equation of motion only affects the metric degree of freedom and in this limiting case Eq. (\ref{motion}) reduce to 
\bea
\frac{\partial \left(\mathcal{S}_G+\mathcal{S}_{G}^{\prime}\right)}{\partial A_{\gamma}}=0.
\eea
Let us then calculate in this scenario the partial derivatives $\frac{\partial \mathcal{S}_G}{\partial A_{\ell}^{\mu}}$ and $\frac{\partial \mathcal{S}_G}{\partial A_{i}^{(0)}}$.
Let us consider the explicit expression of $\mathcal{S}_G$ (\ref{SG})) in terms of the elements of the graph Laplacians, i.e.
\bea
\hspace{-10mm}\mathcal{S}_G=-\Gamma_G\sum_{\mu\neq\nu}(-1)^{\delta_{\mu,t}+\delta_{\nu,t}}\mbox{Tr}\left({\bf L}_{\mu}{\bf L}_{\nu}\right)=-\Gamma_G\sum_{\mu\neq\nu}(-1)^{\delta_{\mu,t}+\delta_{\nu,t}}\sum_{i=1}^N[{\bf L}_{\mu}]_{ii}[{\bf L}_{\nu}]_{ii}.
\eea
Using the explicit expression for the matrix elements $[{\bf L}_{\mu}]_{ii}$ given by Eq.(\ref{Lii}) and considering the link $\ell=[i,j]$ in the direction $\mu$, we obtain
\bea
\frac{1}{4\Gamma_G}\frac{\partial \mathcal{S}_G}{\partial  A_{\ell}^{t}}
=-{[\square-{\bf L}_{t}]_{ii}} e^{A^{t}_{\ell}-A_i^{(0)}}-{[\square-{\bf L}_{t}]_{jj}}e^{A^{t}_{\ell}-A_j^{(0)}},
\eea
and for $\mu\in \{x,y,z\}$,
\bea
\frac{1}{4\Gamma_G}\frac{\partial \mathcal{S}_G}{\partial  A_{\ell}^{\mu}}={[\square+{\bf L}_{\mu}]_{ii}} e^{A^{\mu}_{\ell}-A^{(0)}_i}+{[\square+{\bf L}_{\mu}]_{jj}} e^{A^{\mu}_{\ell}-A_j^{(0)}}.
\eea
Moreover the partial derivative of $\mathcal{S}_G$ with respect to $A^{(0)}_i$ is instead given by
 \bea
\frac{1}{2\Gamma_G}\frac{\partial \mathcal{S}_G}{\partial A_{i}^{(0)}}=\sum_{\mu\neq\nu}(-1)^{\delta_{\mu,t}+\delta_{\nu,t}}[{\bf L}_{\mu}]_{ii}[{\bf L}_{\nu}]_{ii}.
\eea
If $\Gamma\neq 0$ we need to  consider the also the action $\mathcal{S}_{G}^{\prime}$ given by Eq.(\ref{SGp}) which  involves terms of the type $\mbox{Tr}({\bf L}_{\mu}{\bf L}_{\nu}{\bf L}_{\mu}{\bf L}_{\nu})$. The explicit expression of this trace in terms of the elements of the graph Laplacian matrices is given by 
\bea
\hspace*{-15mm}\mbox{Tr}({\bf L}_{\mu}{\bf L}_{\nu}{\bf L}_{\mu}{\bf L}_{\nu})&=&\sum_{i,j,i'j'\in SQ^{\mu\nu}_i}[{\bf L}_{\mu}]_{ij}[{\bf L}_{\nu}]_{ji'}[{\bf L}_{\mu}]_{i'j'}[{\bf L}_{\nu}]_{j'i}
+\sum_{i,j|i\neq j}[{\bf L}_{\mu}]_{ij}[{\bf L}_{\nu}]_{jj}[{\bf L}_{\mu}]_{ji}[{\bf L}_{\nu}]_{ii}\nonumber \\&&+\sum_{i,j|i\neq j}[{\bf L}_{\mu}]_{ii}[{\bf L}_{\nu}]_{ij}[{\bf L}_{\mu}]_{jj}[{\bf L}_{\nu}]_{ji}+\sum_{i}([{\bf L}_{\mu}]_{ii}[{\bf L}_{\nu}]_{ii})^2,
\eea
where the first contribution is a contribution around the squares $SQ^{\mu\nu}_i$ passing through node $i$ and with links in direction $\mu$ and $\nu$ with $\mu\neq\nu$.
From this equation and from the explicit expression of the matrix element of the graph Laplacian in terms of $A^{\mu}_{\ell}$ and $A^{(0)}_i$ (Eq.(\ref{Lii} and Eq. (\ref{Lij})) we can calculate the expression for the partial derivatives
$
\frac{\partial \mathcal{S}_G^{\prime}}{\partial A^{\mu}_{\ell}}$
and 
$
\frac{\partial \mathcal{S}_G^{\prime}}{\partial A^{(0)}_{i}}.$
Considering  the link $\ell=[i,j]$ in direction $\mu$ we obtain
\bea
\hspace*{-15mm}\frac{1}{\Gamma_G^{\prime}}\frac{\partial \mathcal{S}_G^{\prime}}{\partial A^{\mu}_{\ell}}&=&\sum_{\nu\neq\mu}\left\{2\sum_{i'j'\in SQ^{\mu\nu}_{ij}}[{\bf L}_{\mu}]_{ij}[{\bf L}_{\nu}]_{ji'}[{\bf L}_{\mu}]_{i'j'}[{\bf L}_{\nu}]_{j'i}+2[{\bf L}_{\mu}]_{ij}[{\bf L}_{\nu}]_{jj}[{\bf L}_{\mu}]_{ji}[{\bf L}_{\nu}]_{ii}\right.\nonumber \\
&&+4\sum_{j'\neq i}
e^{A_{ij}^{\mu}-A_i^{(0)}}
[{\bf L}_{\nu}]_{ij'}[{\bf L}_{\mu}]_{j'j'}[{\bf L}_{\nu}]_{j'i}
+4\sum_{j'\neq j}
e^{A_{ij}^{\mu}-A_j^{(0)}}
[{\bf L}_{\nu}]_{jj'}[{\bf L}_{\mu}]_{j'j'}[{\bf L}_{\nu}]_{j'j}\nonumber \\
&&\left.+4[{\bf L}_{\mu}]_{ii}([{\bf L}_{\nu}]_{ii})^2
e^{A_{[ij]}^{\mu}-A^{(0)}_i}+4[{\bf L}_{\mu}]_{jj}([{\bf L}_{\nu}]_{jj})^2e^{A_{[ij]}^{\mu}-A^{(0)}_j}\right\}.\eea
Furthermore we obtain
\bea
&&\hspace*{-25mm}\frac{1}{\Gamma_G^{\prime}}\frac{\partial \mathcal{S}_G^{\prime}}{\partial  A^{(0)}_{i}}=-\sum_{\nu\neq\mu}\left\{2\sum_{j,i',j'\in SQ^{\mu\nu}_{i}}\Big([{\bf L}_{\mu}]_{ij}[{\bf L}_{\nu}]_{ji'}[{\bf L}_{\mu}]_{i'j'}[{\bf L}_{\nu}]_{j'i}+[{\bf L}_{\nu}]_{ij}[{\bf L}_{\mu}]_{ji'}[{\bf L}_{\nu}]_{i'j'}[{\bf L}_{\mu}]_{j'i}\Big)\right.\nonumber \\&&\hspace*{-20mm}\left.4\sum_{j\neq i}[{\bf L}_{\mu}]_{ij}[{\bf L}_{\nu}]_{jj}[{\bf L}_{\mu}]_{ji}[{\bf L}_{\nu}]_{ii}+4\sum_{j\neq i}[{\bf L}_{\mu}]_{ii}[{\bf L}_{\nu}]_{ij}[{\bf L}_{\mu}]_{jj}[{\bf L}_{\nu}]_{ji}+4([{\bf L}_{\mu}]_{ii}[{\bf L}_{\nu}]_{ii})^2\right\},
\eea
where $SQ^{\mu\nu}_{ij}$ indicates any of the two plaquettes of direction $\mu\nu$ having link $[i,j]$ as one of their links.\\
If  the matter field $\bm\Psi$ is non-zero we will need to consider the equation of motion  given by Eq. (\ref{motion}) which includes also the  the partial derivatives of the Dirac action $
\frac{\partial \mathcal{S}_D}{\partial A^{\mu}_{\ell}}$ 
and 
$
\frac{\partial \mathcal{S}_D}{\partial A^{(0)}_{i}}$, and the partial derivatives of $\mathcal{S}_R^{\prime}$, $\frac{\partial \mathcal{S}_R^{\prime}}{\partial A^{\mu}_{\ell}}$ 
and 
$
\frac{\partial \mathcal{S}_R^{\prime}}{\partial A^{(0)}_{i}}$.
Let us notice that  the Dirac action $\mathcal{S}_{D}$ can be expressed as 
\bea
\hspace*{-20mm}\mathcal{S}_{D}&=&\bm\chi^{\dag}{\sum_{\mu\in \{x,y,z\}}{\bm \sigma_{\mu}}\otimes\bar{\bf{B}}_{\mu}^{\star}\bm\psi}+\bm\psi^{\dag}\sum_{\mu\in \{x,y,z\}}{\bm \sigma}_{\mu}\otimes\bar{\bf{B}}_{\mu}\bm\chi+\textrm{i}\bm\chi^{\dag}{\bm \sigma}_t\otimes\bar{\bf{B}}_{t}^{\star}\bm\psi+\textrm{i}\bm\psi^{\dag}{\bm\sigma}_t\otimes{\bar{\bf{B}}_{t}\bm\chi}\nonumber \\
\hspace*{-20mm}&&-m({\bm\chi}^{\dag}\bm\chi-\bm\psi^{\dag}\bm\psi).
\eea
Therefore 
\bea
2\frac{\partial \mathcal{S}_{D}}{\partial A_{\ell}^{\mu}}=\left[\bm\chi^{\dag}{\bm \sigma}_{\mu}\otimes\bar{\bf {B}}_{\mu}^{\star}\bm\psi_{\ell}+\bm\psi^{\dag}_{\ell}\sigma_{\mu}\otimes \bar{\bf {B}}_{\mu}\bm\chi\right][1-(1+\textrm{i})\delta_{\mu,t}],
\eea
where $\bm\psi_{\ell}$ is the $2L$ dimensional vector obtained by $\bm\psi$ only keeping the elements $\psi_{[i,j]\pm}$ and $\psi_{[j,i]\pm}$. 
Finally, the derivative of $\mathcal{S}_D$ with respect to $A_i^{(0)}$ is given by 
%\begin{widetext}
\bea
\hspace{-25mm}-2\frac{\partial \mathcal{S}_{D}}{\partial A_{i}^{(0)}}=\bm\chi^{\dag}_i\left[\sum_{\mu\in \{x,y,z\}}{\bm \sigma}_{\mu}\otimes \bar{\bf{B}}_{\mu}^{\star}+\textrm{i}{\bm \sigma}_t\otimes\bar{\bf{B}}_{t}^{\star}\right]\bm\psi+\bm\psi^{\dag}\left[\sum_{\mu\in \{x,y,z\}}{\bm \sigma}_{\mu}\otimes \bar{\bf{B}}_{\mu}+\textrm{i}{\bm\sigma}_t\otimes\bar{\bf{B}}_{t}\right]\bm\chi_i,\nonumber 
\eea
%\end{widetext}
where $\bm\chi_i$ is the $2N$ dimensional vector obtained from $\bm\chi$ by considering only the elements corresponding to node $i$.\\
The action $\mathcal{S}_R^{\prime}$ can be  by expressed as 
\bea
\mathcal{S}_R^{\prime}=\bar{\bm \Psi}\sum_{\mu,\nu}[-\Gamma_C F_{\mu\nu}+\Gamma_A G_{\mu\nu}]{\bm\Psi+}={\bm \psi}^{\dag}\sum_{\mu,\nu}g_{\mu\nu}{\bm\psi}.
\eea
where $g_{\mu\nu}=g_{\nu\mu}$ is given by 
\bea
g_{\mu\nu}=\Gamma_C \epsilon_{\mu\nu\theta}\left[\bm\sigma^{\theta}\otimes (\bar{\bf B}_{\mu}{\bar{\bf B}}_{\nu}^{\star}+{\bar{\bf B}}_{\nu}{\bar{\bf B}}_{\mu}^{\star})\right]
\eea
for $\mu,\nu\in \{x,y,z\}$ and $g_{t\mu}$ is given by 
\bea
g_{t\mu}=\Gamma_A \bm\sigma_\mu\otimes (\bar{\bf B}_{t}{\bar{\bf B}}_{\mu}^{\star}+{\bar{\bf B}}_{\mu}{\bar{\bf B}}_{t}^{\star}).
\eea
The partial derivative of $\mathcal{S}_{R}^{\prime}$ with respect to the link metric $A_{\ell}^{\mu}$ is given by 
\bea
{2}\frac{\partial \mathcal{S}_{R}^{\prime}}{\partial A_{\ell}^{\mu}}={\bm \psi}^{\dag}_{\ell}\sum_{\mu,\nu}g_{\mu\nu}{\bm\psi}+{\bm \psi}^{\dag}\sum_{\mu,\nu}g_{\mu\nu}{\bm\psi}_{\ell},
\eea
for $\mu\in \{t,x,y,z\}$. The partial derivative of $\mathcal{S}_{R}^{\prime}$ with respect to the link metric $\partial A_{i}^{(0)}$ is given by
\bea
-\frac{\partial \mathcal{S}_{R}^{\prime}}{\partial A_{i}^{(0)}}={\bm \psi}^{\dag}\sum_{\mu,\nu}g_{\mu\nu i}{\bm\psi},
\eea
where $g_{\mu\nu i}$ is given by 
\bea
g_{\mu\nu i}=\Gamma_C \epsilon_{\mu\nu\theta}\left[\bm\sigma^{\theta}\otimes (\bar{\bf B}_{\mu}{\bf e}_i{\bf e}_i^{\top}{\bar{\bf B}}_{\nu}^{\star}+{\bar{\bf B}}_{\nu}{\bf e}_i{\bf e}_i^{\top}{\bar{\bf B}}_{\mu}^{\star})\right]
\eea
for $\mu,\nu\in \{x,y,z\}$ and $g_{t\mu}$ is given by 
\bea
g_{t\mu}=\Gamma_A \bm\sigma_\mu\otimes (\bar{\bf B}_{t}{\bf e}_i{\bf e}_i^{\top}{\bar{\bf B}}_{\mu}^{\star}+{\bar{\bf B}}_{\mu}{\bf e}_i{\bf e}_i^{\top}{\bar{\bf B}}_{t}^{\star}),
\eea
with ${\bf e}_i$ indicating the topological spinor whose only non-zero elements are the ones corresponding to node $i$.\\
We conclude this section noting  that in the different limiting case in which ${\bf G}_0$ and ${\bf G}_1$ are diagonal but ${\bf G}_{[1]}^{-1}([i,j],[i,j])=-{\bf G}_{[1]}^{-1}([j,i],[j,i])$ then we would have in general $\mathcal{S}_R=R\neq 0$ while $\mathcal{S}_G=\mathcal{S}_{G}^{\prime}=0.$
\section{Conclusions}
In conclusion in this work we are proposing a gauge theory where the matter fields are defined on both the nodes and the links of the network, similar to the classical situation in which the state of a particle is specified by both position and velocity.
The matter field can be treated by an action making use of the discrete Dirac operator. This operator, defined in Ref. \cite{bianconi2021topological} for $3+1$ dimensional lattices is here extended to treat directed and weighted links. In particular we consider the case in which the network is associated with metric matrices ${\bf G}_{[0]}$ and ${\bf G}_{[1]}$ as it is usual in algebraic topology.
 However while in applied topology these matrices are typically considered diagonal, here we assume that in general they are not diagonal and that their matrix elements constitute the geometrical degree of freedom associated to the network.
Our work shows that the these metric matrices are determining  the electromagnetic field associated to this theory. Indeed the non-relativistic limit of the Dirac equation confirms our interpretation  of  the anticommutator of the spatial directional Dirac operators as the magnetic field.  
Moreover the non-relativistic limit of the Dirac equation allows us to draw the following important additional conclusions: in this limit, the wave function of the electron defined on the links (vector bundle) follows the Schr\"odinger equation of the electron, with the correct giromagnetic moment; the wave function defined on the nodes, instead follows the Klein-Gordon equation and it is not negligible. These findings  might lead to observable differences between the non-relativistic limit of the traditional Dirac equation and possibly might be related to super-symmetric models.
We then consider the  gauge transformation acting on the Dirac spinor and the metric matrices.
We then consider the specific case in which the  Dirac topological fields interacts with the underlying degrees of freedom of the network geometry.
The action associated to the metric field can be constructed contracting the curvature tensor of our theory, which is defined as the non-vanishing commutator of the directional Dirac operators associated to different directions.
The resulting gauge transformations of metric and matter fields are discussed and the equations of motion are derived  in the specific example in which the metric fields are real and diagonal.

This works can be extended in different directions.
First of all our approach has focused only on networks, i.e. the topological spinor considered in this work is only defined on nodes and links. However it would be interesting to explore in the future whether this framework can be extended also to topological spinors defined on the higher-order cells of the lattice such as the squares, and the cubes and if this extension will bring new physical results.

Secondly, the approach could be possibly generalized to treat  invariance under diffeomorphisms. 

Thirdly, we have  restricted our analysis to  an underlying network topology formed by a  $3+1$ dimensional lattice  but this framework can be applied easily also to lattices of dimension $1+1$ and $2+1$.  The choice of a  $3+1$ square  lattice   can be considered as a first approximation of a general network topology  valid for  almost flat spaces. The underlying simple lattice topology allows us to adopt the  Dirac operator enriched by the Lie algebra of Pauli matrices for distinguishing between exterior derivatives in different directions as proposed in Ref.\cite{bianconi2021topological}. However on more general topologies other Lie algebras should be adopted.

Finally the approach here developed for the Dirac equation can be potentially extended to treat Majorana fermions \cite{majorana1937teoria,wilczek2009majorana}.

We hope that this work will stimulate further discussion on these open questions and more in general  in the emergent field of gauge theories on networks which is both relevant for different approaches to quantum gravity and for realization of artificial  gauge theories in condensed matter with applications in quantum computing.
  
After the submission of this work, the proposed operator has been adopted by Shahn Majid for formulating a non-commutative spectral triple in Ref. \cite{majid2023dirac} providing further results that nicely comment our discussion. 

\appendix
\section{Choice of the Hodge-Dirac operator $d_{\mu}+d_{\mu}^{\star}$}
In this work we have defined the Hodge-Dirac operator $d_{\mu}+d_{\mu}^{
\star}$ as the self-adjoint operator
\bea
d_{\mu}+d^{\star}_{\mu}=\left(\begin{array}{cc}
0&\bm \sigma_0\otimes{\bf \bar{B}}_{\mu}^{\star}\\
\bm\sigma_0\otimes {\bf \bar{B}}_{\mu}&0
\end{array}\right),
\label{ext_muw} 
\eea
where ${\bf \bar{B}}_{\mu}^{\star}={\bf \bar{B}}_{\mu}^{\dag}$ is adjoint to ${\bf \bar{B}}_{\mu}$ according to the $L^2$ norm between $0$-cochains and $1$-cochains defined in Eq.(\ref{scalar_01}).
In this Appendix we point out  that it is very straightforward to check that  the considered Hodge-Dirac operator $d_{\mu}+d^{\star}_{\mu}$   is isospectral to the asymmetric Hodge-Dirac operator (see \cite{baccini2022weighted} for details) defined as 
\bea
\left(\begin{array}{cc}
0&\bm \sigma_0\otimes[{\bf \bar{B}}^{(U)}_{\mu}]^{\star}\\
\bm\sigma_0\otimes {\bf \bar{B}}^{(U)}_{\mu}&0
\end{array}\right),
\label{ext_muw2} 
\eea
where 
\bea
[{\bf \bar{B}}^{(U)}_{\mu}]^{\star}={\bf G}_{[0]}[{\bf \bar{B}}^{(U)}_{\mu}]^{\dag}{\bf G}_{[1]}^{-1}
\eea
is the adjoint operator of the unweighted coboundary operator ${\bf \bar{B}}^{(U)}_{\mu}$ with respect to the (weighted norms) between $0$-cochains ($\hat{\bm \chi},{\bm \chi}$) and $1$-cochains ($\hat{\bm \psi},{\bm \psi}$)  defined as
\bea
(\hat{\bm \chi},{\bm \chi})=\hat{\bm \chi}^{\dag}{\bf G}_{[0]}^{-1}{\bm \chi},\quad
(\hat{\bm \psi},{\bm \psi})=\hat{\bm \psi}^{\dag}{\bf G}_{[1]}^{-1}{\bm \psi}.
\eea

\section*{Acknowledgments}
G. Bianconi acknowledges interesting discussions with  Marcus Reitz, and Shahn Majid and 2021 conversations with Juergen Jost on a previous version of this manuscript.
\section*{References}
\bibliographystyle{unsrt}
\bibliography{references}
\end{document}